\documentclass[aps,pra,showpacs,groupedaddress,twocolumn,amsmath,amssymb,10pt]{revtex4-1}

\usepackage[american]{babel}
\usepackage{graphicx}
\usepackage{times}
\usepackage{color}
\usepackage{grffile}
\newcommand{\ket}[1]{| #1 \rangle}
\newcommand{\bra}[1]{\langle #1 |}
\newcommand{\w}{{\rm{\textit w}}}

\begin{document}
\title{Universal nonclassicality witnesses for harmonic oscillators}
\author{T. Kiesel and W. Vogel}

\affiliation{Arbeitsgruppe Quantenoptik, Institut f\"ur Physik, Universit\"at  Rostock, D-18051 Rostock,
Germany}
\begin{abstract}
It is shown that a nonclassicality witness, whose expectation value can be measured for all quantum states, can be constructed from every nonclassicality filter. This finding leads to a set of universal witnesses,  parameterized by only three real numbers, for the detection of nonclassicality of any quantum state of a harmonic oscillator. An explicit operator expression is given for such a universal witness. The application of the witnesses is demonstrated for a nontrivial example, and its experimental measurement is briefly considered.
\end{abstract}

\maketitle

\section{Introduction}
The notion of nonclassicality describes the difference between classical and quantum physics. For the harmonic oscillator, its formal definition is based on the coherent states $\ket\alpha$, which are  the closest analogues to the classical oscillation~\cite{Schrodinger}. Any quantum state $\hat \rho$ can be formally written as a statistical mixture of (classical) coherent states~\cite{Sudarshan,Glauber},
\begin{equation}
	\hat \rho = \int d^2\alpha\,P(\alpha)\ket\alpha\bra\alpha.
\end{equation}
The function $P(\alpha)$ is the so-called Glauber Sudarshan $P$~function. 
If it is a classical probability density, the state $\hat \rho$ is a classical mixture of coherent states. In many cases, however, the $P$~function is a  quasiprobability, violating properties of a probability. Then the state is called nonclassical~\cite{TitulaerGlauber}. 

In general this definition cannot be applied in most practical situations, since the $P$~function   may have strong singularities. Therefore, several criteria for nonclassicality have been developed,
which can be distinguished in two major categories. First, one can examine the expectation value of a so-called nonclassicality witness $\hat W$ for a given  quantum state~\cite{Shchukin05,Korbicz05}. Without loss of generality, this operator has a nonnegative expectation value for all classical states $\hat{\rho}_{\rm cl}$,
\begin{equation}
{\rm Tr}\{\hat{\rho}_{\rm cl} \hat{W}\} \geq 0.
\end{equation}
Conversely, if this inequality is violated for 
a quantum state $\hat{\rho}$, 
then this state 
must be a nonclassical one:
\begin{equation}
	{\rm Tr}\{\hat{\rho}\hat{W}\} < 0\quad\Rightarrow\quad \hat{\rho}\ \mbox{nonclassical}.
\end{equation}
 Based on nonclassicality witnesses, one may formulate conditions for matrices of characteristic functions~\cite{Ri-Vo}, moments~\cite{Shchukin05b,Miranowicz10}, or outcome probabilities~\cite{Luis79}. 

On the other hand, one may examine phase-space representations of a quantum state, which play the role of a probability density in some sense. Besides the $P$ function, the Wigner function is frequently used in experiments since it is always regular.
For nonclassical states, the Wigner function may 
become negative for some points in phase space. 
These negativities indicate the nonclassicality of the quantum state. 
However, nonnegativity of the Wigner function does not prove classicality, and 
negative values of the Wigner function are only sufficient for nonclassicality.
The recently developed concept of nonclassicality filters and  quasiprobabilities provides a complete characterization of nonclassical effects~\cite{Kiesel10} in terms of regular phase-space distributions.

In the present paper we show that generally applicable nonclassicality witnesses can be constructed from any nonclassicality filter. In Sec. II, we elaborate the connection to the previously known nonclassicality witnesses and provide an explicit analytical expression of a universal witness. Experimental applications are briefly discussed in Sec.~III. In Sec. IV, the usefulness of our method is illustrated for  an example of a quantum state, under conditions when many other tests fail.

\section{Universal nonclassicality witnesses}
\subsection{Nonclassicality filters and witnesses}
Let us start from the characteristic function of the $P$~function of the state 
$\hat \rho$,
\begin{equation}
	\Phi(\beta) = {\rm Tr}\left\{\hat\rho e^{\beta\hat a^\dagger}e^{-\beta^*\hat a}\right\}. \label{eq:def:Phi}                                                                                   
\end{equation}
We choose a function of a complex argument $\beta$, $\Omega_\w(\beta)=\Omega_1(\beta/\w)$, satisfying the conditions~\cite{Kiesel10}:
\begin{enumerate}
	\item[C1.] $\Omega_\w(\beta) e^{|\beta|^2/2}$ is integrable for all positive $\w$.
	\item[C2.] $\Omega_\w(\beta)$ has a nonnegative Fourier transform.
	\item[C3.] $\Omega_\w(0) = 1$ and $\lim_{\w\to\infty} \Omega_\w(\beta) = 1$, for all $\beta$.
\end{enumerate}
Then, $\Omega_\w(\beta)$ is a family of nonclassicality filters, parameterized by the real number $\w$. 

The Fourier transform of the filtered characteristic function, 
\begin{equation}
  P_\w(\alpha) = \frac{1}{\pi^2} \int d^2\beta\, \Phi(\beta) \Omega_{\w}(\beta) e^{\alpha\beta^*-\alpha^*\beta}.
  \label{eq:def:P:Omega}
\end{equation}
is a filtered $P$ function of the state. Since the filter is designed for the requirements of nonclassicality tests, $P_\w(\alpha)$ is denoted as nonclassicality-filtered $P$ function (NFP). 
The NFP is nonnegative for all classical states.  For all nonclassical states, its negativities uncover nonclassicality for properly chosen $\w$ and $\alpha$. When the NFPs contain the full information about the quantum state,
they are called nonclassicality quasiprobabilities~\cite{Kiesel10}. 
Their experimental reconstruction has been demonstrated~\cite{Kiesel11-1,Kiesel11-2}.
In the following we renounce this additional constraint and consider general NFPs.

\subsection{Construction of nonclassicality witnesses}
In the following, we rewrite the NFP as the expectation value of an Hermitian operator with finite trace. For this purpose, we insert Eq.~(\ref{eq:def:Phi}) into the definition of the NFP~(\ref{eq:def:P:Omega}) and obtain
\begin{equation}
  P_\w(\alpha) = {\rm Tr}\left\{\hat \rho \hat W_\w(\alpha)\right\} = \langle\hat W_\w(\alpha)\rangle,\label{eq:link:P_Omega:and:witness}
\end{equation}
with the operator 
\begin{equation}
  \hat W_\w(\alpha) = \frac{1}{\pi^2}\int d^2\beta \, \Omega_{\w}(\beta)  e^{\beta(\hat a^\dagger-\alpha^*)} e^{-\beta^*(\hat a-\alpha)}.\label{eq:def:witness}
\end{equation}
The complex argument $\alpha$ plays the role of a coherent displacement of the operator $\hat W_\w \equiv \hat W_\w(0)$, it can be expressed with the unitary displacement operator $\hat D(\alpha) = e^{\alpha a^\dagger-\alpha^*\hat a}$ by
\begin{equation}
  \hat W_\w(\alpha) = \hat D(\alpha) \hat W_\w \hat D(-\alpha).\label{eq:def:witness:by:D} 
\end{equation}
Let us consider the operator $\hat W_\w$ first. Because $\Omega_\w(\beta)$ has a real Fourier transform, we have $\Omega_\w(-\beta) = \Omega^*_\w(\beta)$, which implies that $\hat W_\w$ is Hermitian. Furthermore, the expectation value for any coherent state is given by the Fourier transform of $\Omega_\w(\beta)$,
\begin{equation}
  \bra\alpha\hat W_\w\ket\alpha = \frac{1}{\pi^2}\int d^2\beta \, \Omega_{\w}(\beta)  e^{\beta \alpha^* -\beta^*\alpha}, \label{eq:Fourier:trafo:Omega}
\end{equation}
which is nonnegative by definition of a nonclassicality filter. In consequence, the same holds not only for coherent states, but for all classical states. Last, $\hat W_\w(\alpha)$ is a trace class operator, since ${\rm Tr}\{e^{\beta\hat a^\dagger}e^{-\beta^*\hat a}\} = \pi\delta(\beta)$ and $\Omega_\w(0) = 1$:
\begin{equation}
	{\rm Tr}\{\hat W_\w\} = \frac{1}{\pi^2}\int d^2\beta\, \Omega_\w(\beta)
			{\rm Tr}\{e^{\beta\hat a^\dagger}e^{-\beta^*\hat a}\}= \frac{1}{\pi}. 
\end{equation}
Since the coherent displacement does not affect these properties, we conclude that $W_\w(\alpha)$ has the same features for all~$\alpha$. 

This gives rise to the interpretation of the Hermitian operator $\hat W_\w(\alpha)$ as a \textit{nonclassicality witness}: Its expectation value is nonnegative for all classical states. Hence, states with negative expectation values are clearly identified as being nonclassical. In this sense, we have proved that any nonclassicality filter $\Omega_\w(\beta)$ can be used to define a nonclassical witness $\hat W_\w(\alpha)$ by Eq.~(\ref{eq:def:witness}). Its expectation value is the NFP $P_\w(\alpha)$, see Eq.~(\ref{eq:link:P_Omega:and:witness}). Finally, by adapting the results in~\cite{Kiesel10}, we know that for any nonclassical state and any nonclassicality witness $\hat W_\w(\alpha)$ as defined above, there exists a real width parameter $\w$ and a complex number $\alpha$ such that the expectation value of $\hat W_\w(\alpha)$ is negative. Therefore, we constructed a set of nonclassicality witnesses, parameterized by only three real numbers, which is sufficient for the detection of nonclassicality of an arbitrary quantum state. In this sense, any witness constructed in this way is universal.

In order to demonstrate the equivalence of witnesses and filters, we still have to show the converse statement: Whenever one has a nonclassicality witness $\hat W$ with ${\rm Tr}(\hat W) = \frac{1}{\pi}$, as will be assumed in the following,  one may construct a nonclassicality filter from it. For this purpose, let us represent the witness $\hat W$ by its characteristic function $\Phi^{(Q)}_{\hat W}(\beta)$ of the $Q$ function~\cite{Perelomov},
\begin{equation}
	\hat W = \frac{1}{\pi^2}\int d^2\beta \, \Phi^{(Q)}_{\hat W}(\beta) e^{\beta\hat a^\dagger} e^{-\beta^*\hat a}.\label{eq:def:witness:from:PhiQ}
\end{equation}
Now, one easily sees from Eq.~\eqref{eq:def:witness} that the characteristic function can be directly connected to a nonclassicality filter:
\begin{equation}
	\Omega_\w(\beta) = \Phi^{(Q)}_{\hat W}(\beta),\label{eq:relate:Omega:to:Phi}
\end{equation}
and the corresponding NFP is given by Eq.~\eqref{eq:link:P_Omega:and:witness}. Because of its bounded trace, the expectation value of $\hat W$ is always finite. Hence the NFP is always a regular function, and may be accessible to experiments. Still, the width parameter $\w$ is not implemented, since the right hand side of~Eq.~\eqref{eq:relate:Omega:to:Phi} does not depend on a real parameter $\w$. However, in Appendix A we show how this can be achieved.

\subsection{Explicit form of universal witness operators}\label{sec:explicit:filter}

Let us consider the nonnegativity of the Fourier transform of the filter $\Omega_\w(\beta)$ more in detail. This condition can by written with the help of Eq.~\eqref{eq:Fourier:trafo:Omega} as
\begin{equation}
	\bra\alpha\hat W_\w\ket\alpha = \tilde \omega_\w^*(\alpha^*,\alpha)\tilde\omega_\w(\alpha^*,\alpha),\label{eq:nonnegativity}
\end{equation}
with some suitable function $\tilde\omega_\w(\alpha^*,\alpha)$. Here, we write the arguments $\alpha$ and $\alpha^*$ explicitly in order to have a clear notation of corresponding operators which will be defined below. The nonclassicality filter is the autocorrelation function 
\begin{equation}
	\Omega_\w(\beta) = \int d^2\beta'\,\omega_\w(\beta')\omega_\w\left(\beta+\beta'\right)  \label{eq:autocorrelation:filter}
\end{equation}
of the Fourier transform of $\tilde\omega_\w(\alpha^*,\alpha)$,
\begin{equation}
	\omega_\w(\beta) = \frac{1}{\pi} \int d^2\alpha\, \tilde\omega_\w(\alpha^*,\alpha) e^{\alpha^*\beta-\alpha\beta^*}.\label{eq:Fourier:tilde:omega}
\end{equation}
Note that $\Omega_\w(0) = 1$ sets a normalization condition on $\tilde\omega_\w(\alpha^*,\alpha)$. However, it is not needed for nonclassicality tests, but only for the correct normalization of quasiprobabilities. Furthermore, since we consider $\Omega_\w(\beta) = \Omega_1(\beta/\w)$, the width parameter enters into $\tilde\omega(\beta)$ as
\begin{equation}
	\tilde\omega_\w(\alpha^*,\alpha) = \w\,\tilde\omega_1(\w\alpha^*,\w\alpha).\label{eq:scale:tilde:omega}
\end{equation}
Most importantly, $\tilde\omega(\alpha^*,\alpha)$ has to be chosen such that the function $\Omega_\w(\beta)$ still fulfills condition C1. It has been shown to be sufficient that $\omega_\w(\beta)$ itself fulfils this requirement~\cite{Kiesel10}, i.e.~$\omega_\w(\beta)e^{|\beta|^2/2}$ is integrable for all $\w$.

From Eq.~\eqref{eq:nonnegativity}, we easily find the normally-ordered form of the nonclassicality witness $\hat W_\w$. It is obtained by replacing the c-numbers $\alpha,\alpha^*$ by the operators $\hat a,\hat a^\dagger$:
\begin{equation}
	\hat W_\w = :\!\tilde\omega_\w^\dagger(\hat a^\dagger,\hat a) \tilde\omega_\w(\hat a^\dagger,\hat a):,
\end{equation}
the notation $:\,:$ indicates that all products of the operators $\hat a^\dagger,\hat a$ have to be written in normal order. Furthermore, by defining the displaced operator
\begin{equation}
	\tilde\omega_{\w,\alpha}(\hat a^\dagger,\hat a) = \tilde\omega_\w(\hat a^\dagger-\alpha^*, \hat a-\alpha),
	\label{eq:displace:tilde:omega}
\end{equation}
the relation~\eqref{eq:def:witness:by:D} leads us to 
\begin{equation}
	\hat W_\w(\alpha) = :\!\tilde\omega_{\w,\alpha}^\dagger(\hat a^\dagger,\hat a) \tilde\omega_{\w,\alpha}(\hat a^\dagger,\hat a):.\label{eq:witness:omega:2}
\end{equation}
This is a remarkable result. First, we can construct nonclassicality witnesses easily by
properly choosing the function $\tilde\omega_\w(\alpha^*,\alpha)$. It has to satisfy some weak conditions, in particular its Fourier transform obeys
the condition C1. The other two conditions are then trivially satisfied by construction.
Second, it states that any nonclassicality witness, which can be constructed by a nonclassicality filter, can be written in such a form. The operator $\tilde\omega_{\w,\alpha}(\hat a^\dagger,\hat a)$ can be found from Eqs.~\eqref{eq:Fourier:trafo:Omega}, \eqref{eq:nonnegativity}, \eqref{eq:displace:tilde:omega}, and only depends on three real parameters. As already discussed, one particular operator $\tilde\omega_{\w,\alpha}(\hat a^\dagger,\hat a)$, equipped with these three parameters, is sufficient to completely determine nonclassicality of an arbitrary state.

\subsection{Example of a universal witness} 
Here we consider a witness that can be given explicitly. We fix $\tilde\omega_\w(\hat a^\dagger,\hat a)$ by choosing its Fourier transform as the disc function
\begin{equation}
	\omega_1(\beta) = \left\{\begin{array}{l l} 1, & |\beta| < 1/2,\\ 0, & \mbox{elsewhere.} \end{array}\right.
\end{equation}
The corresponding nonclassicality filter does not preserve all information about a quantum state, since it has bounded support. Furthermore, it is not appropriately normalized. However, it is suitable for the construction of a universal set of nonclassicality witnesses. 

From the inverse relation of~\eqref{eq:Fourier:tilde:omega} and Eqs.~\eqref{eq:scale:tilde:omega}, \eqref{eq:displace:tilde:omega}, \eqref{eq:witness:omega:2}, we find the nonclassicality witness:
\begin{equation}
	\hat W_\w(\alpha) = :\frac{[J_1(\w \sqrt{(\hat a^\dagger-\alpha^*)(\hat a-\alpha)})]^2}{4(\hat a^\dagger-\alpha^*)(\hat a-\alpha)}: ,
\label{witn-expl}
\end{equation} 
where $J_1(x)$ is the Bessel function of first order. Equation~\eqref{witn-expl} represents a complete family of witness operators in a closed analytic form, which is
an important finding of our paper. It is suited to identify all the nonclassical effects of any quantum state of a harmonic oscillator. This expression is well-defined in general. It can be expanded into a normally ordered power series of the displaced photon number operator $\hat n(\alpha) = (\hat a^\dagger-\alpha^*)(\hat a-\alpha)$,
\begin{equation}
	\hat W_\w(\alpha) =  \frac{\w^2}{16} \sum_{m=0}^\infty \frac{(-\w^2/4)^m}{[(m+1)!]^2}\binom{2m+2}{m} :\hat n(\alpha)^m\!:.\label{eq:triangular:witness}
\end{equation}
It is noteworthy that this form of the witness is not unique. As we have shown, different witnesses can be constructed from different nonclassicality filters, defined by different functions $\omega_1(\beta)$. All these witnesses are equivalent for the task to uncover nonclassical effects completely. 

\subsection{Relation to other witnesses}
Nonclassicality witnesses of a form similar to~\eqref{eq:witness:omega:2} have already been examined in~\cite{Shchukin05,Korbicz05,Shchukin05b}. They have been constructed as normally ordered squares of some operator $\hat f = \hat f(\hat a^\dagger,\hat a)$,
\begin{equation} 
	\hat W = :\!\!\hat f^\dagger(\hat a^\dagger,\hat a)\hat f(\hat a^\dagger,\hat a)\!\!:.\label{eq:witness:f2}
\end{equation}
Using the $P$ function, its expectation value is given by
\begin{equation}
   \langle:\!\!\hat f^\dagger(\hat a^\dagger,\hat a)\hat f(\hat a^\dagger,\hat a)\!\!:\rangle = \int P(\alpha) |f(\alpha^*,\alpha)| d^2\alpha.\label{eq:expect:witness:f2}
\end{equation}
Clearly, negative expectation values can only be achieved by negativities of the $P$ function, i.e.~by nonclassical states. Simple examples lead to well known nonclassicality conditions.
Choosing $\hat f = \hat x - \langle\hat x\rangle$, where $\hat x$ is the quadrature operator, we may identify squeezing~\cite{Squeezing}. For $\hat f = \hat n - \langle\hat n\rangle$, $\hat n = \hat a^\dagger \hat a$, we uncover sub-Poissonian photon number statistics~\cite{SubPoisson}.

More general nonclassicality tests can be obtained from Eq.~(\ref{eq:witness:f2}).
One may derive a complete hierarchy of conditions in terms of characteristic functions~\cite{Ri-Vo},
which directly follows via Fourier representation of the operator $\hat f(\hat a^\dagger,\hat a)$, cf.~\cite{Shchukin05b}. Via Taylor series expansion of $\hat f(\hat a^\dagger,\hat a)$
one obtains criteria in terms of matrices of moments~\cite{Shchukin05,Shchukin05b,Miranowicz10}. 
The former method can be applied to any quantum state. In general, however, 
such a nonclassicality test may become rather complex.
The latter method is only applicable provided that the needed moments exist.
It also becomes complex, in particular if large matrices are needed to detect the nonclassical effects. 
If all moments of a quantum state exist, the test operators $\hat f$ can be restricted to polynomials in $\hat a,\hat a^\dagger$~\cite{Korbicz05}, which, however, still form an infinite dimensional space.

In contrast, our approach is much simpler, since it requires only control of three real parameters, in order to perform complete nonclassicality tests.  They enter into the operator $\tilde\omega_{\w,\alpha}(\hat a^\dagger,\hat a)$ simply as a displacement and rescaling.
On the other hand, our method does not rely on the assumption that all moments of the quantum state exist. This is due to the requirement C1 for the nonclassicality filter, which guarantees the existence of the quasiprobabilities and expectation values of the witnesses. Remarkably, this condition is much more fundamental as it might look at first sight. It is also necessary for a complete test of nonclassicality of states, for which \textit{all} moments exist. For instance, if we choose $\hat f_{\w,\alpha}(\hat a^\dagger,\hat a) = \w(\hat a-\alpha)$, it has the same structure as our $\tilde\omega_{\w,\alpha}(\hat a^\dagger,\hat a)$ in Eq.~\eqref{eq:displace:tilde:omega}, but cannot be used to construct a nonclassicality filter which fulfils the condition C1. As a witness, with $\hat f= \hat f_{\w,\alpha}$ in Eq.~(\ref{eq:witness:f2}), it requires the existence of the first and second moments of $\hat a^\dagger$, $\hat a$.
However, this does not provide a nonclassicality test, even for a single quantum state!

\section{Direct measurement of nonclassicality witnesses}

From Eq.~(\ref{eq:link:P_Omega:and:witness}), it is clear that one may calculate the expectation value of the witness $\hat W_\w(\alpha)$ by reconstruction of the NFP $P_\w(\alpha)$, analogously to the work in~\cite{Kiesel11-1,Kiesel11-2}. However, this procedure requires full quantum tomography of the state. The question arises if one can estimate the expectation value directly. 

Let us restrict to the case of phase-independent nonclassicality filters, i.e.~$\Omega_\w(\beta) = \Omega_\w(|\beta|)$. The resulting nonclassicality witness $\hat W_\w$ is diagonal in photon number basis, cf. e.g. Eq.~(\ref{eq:triangular:witness}). Its expectation value can be written as
\begin{equation}
	\langle\hat W_{\w}\rangle = \sum_{n=0}^\infty \bra n\hat W_\w\ket n p_n,\label{eq:witness:from:probabilities}
\end{equation}
with $p_n$ being the photon number statistics of the quantum state. Since the  latter can be measured by photon-number resolving detectors~\cite{Raymer}, and the matrix elements of $\hat W_\w$ can be calculated theoretically for arbitrary $\w$, one can determine this expectation value with standard experimental techniques. 

Of course, practically one is restricted to a finite number of photons, 
approximating the series in Eq.~\eqref{eq:witness:from:probabilities} by a finite sum. The smaller the width parameter, the better the approximation will be, since the matrix elements $\bra n\hat W_\w\ket n$ are polynomials in $\w^2$, cf.  Eq.~\eqref{eq:triangular:witness}. For larger $\w$~parameter, a larger number of photons 
must be resolved, which can be difficult in practice.
This problem is similar to the practical limitation of the $\w$~parameter as discussed in the context of nonclassicality quasiprobabilities~\cite{Kiesel11-1}. In such cases one may estimate a systematic error due to the truncation
of the series~\eqref{eq:witness:from:probabilities}.

\begin{figure}
   \includegraphics[width=0.7\columnwidth]{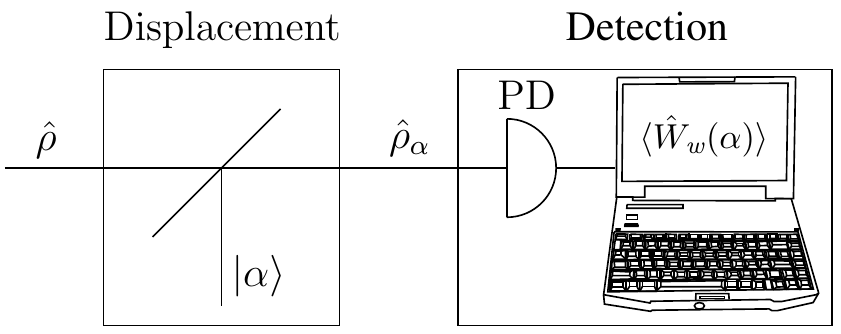}
   \caption{Experimental setup. The beamsplitter displaces the initial state, followed by a 
photon-number resolving detection. The filter width is introduced by application of Eq.~\eqref{eq:expect:in:Fock:basis}.}
   \label{fig:BS}
\end{figure}

In order to measure the expectation value of $\hat W_\w(\alpha)$, we still have to include the coherent displacement in Eq.~\eqref{eq:def:witness:by:D}. For this purpose, it is indifferent if we displace the witness operator by the amplitude $\alpha$ or the quantum state by the amplitude $-\alpha$,
\begin{equation}
\langle \hat W_\w(\alpha) \rangle
= \sum_{n = 0}^\infty  \bra n\hat W_\w \ket n  p_n(-\alpha),
  \label{eq:expect:in:Fock:basis}
\end{equation}
with $p_n(\alpha) = \bra n\hat\rho_\alpha\ket n$ and $\hat\rho_\alpha = \hat D(-\alpha)\hat\rho\hat D(\alpha)$.
This can be done by methods considered in~\cite{Kis99,Raymer,Wallentowitz96}: a highly transparent beamsplitter is used to  displace the original state by a coherent one and a photon-number resolving detector measures the signal, see Fig.~\ref{fig:BS}. 
Then, the expectation value of the phase-independent nonclassicality witness $\hat W_\w$ is determined for the displaced state, which equals to the expectation value of $\hat W_\w(\alpha)$ for the original state.
In this way, one can choose $\alpha$ by the experimental setup, while the width parameter is included in the Fock matrix elements of $\hat W_\w$. The variation of these three real parameters enables one to perform a complete nonclassicality test for an arbitrary quantum state. 
Note that the practical determination of $\langle \hat W_\w(\alpha) \rangle$ can also be based on
direct sampling of the photon number statistics from phase-randomized quadrature measurements, as done in~\cite{Raymer}.

\section{Application}

Let us consider a single photon added thermal state 
(SPATS)~\cite{AgarwalTara}, characterized by the mean photon number of 
the thermal background, $\bar n$, and a quantum efficiency $\eta$. It is 
known that for $\bar n\geq 1/\sqrt{2}$, this state does exhibit neither 
squeezing nor sub-Poissonian statistics, arising from 
Eq.~(\ref{eq:witness:f2}) with $\hat f = \hat x - \langle\hat x\rangle$ and 
$\hat f = \hat n - \langle\hat n\rangle$ respectively~\cite{Shchukin05}. 
Furthermore, for $\bar n \geq 0.39$, the characteristic function does not show nonclassicality of first order~\cite{Ri-Vo}. Therefore, these well-known nonclassicality witnesses are not able 
to verify nonclassicality of this particular state.
It is also important that for $\eta \leq 1/2$, the  Wigner function is 
positive semidefinite~\cite{Bellini}.
Hence, by setting the quantum efficiency $\eta = 1/2$ and choosing $\bar n\geq 1/\sqrt{2}$, it is a 
nontrivial task to witness the nonclassicality of the state under study.

\begin{figure}
   \includegraphics[width=0.9\columnwidth]{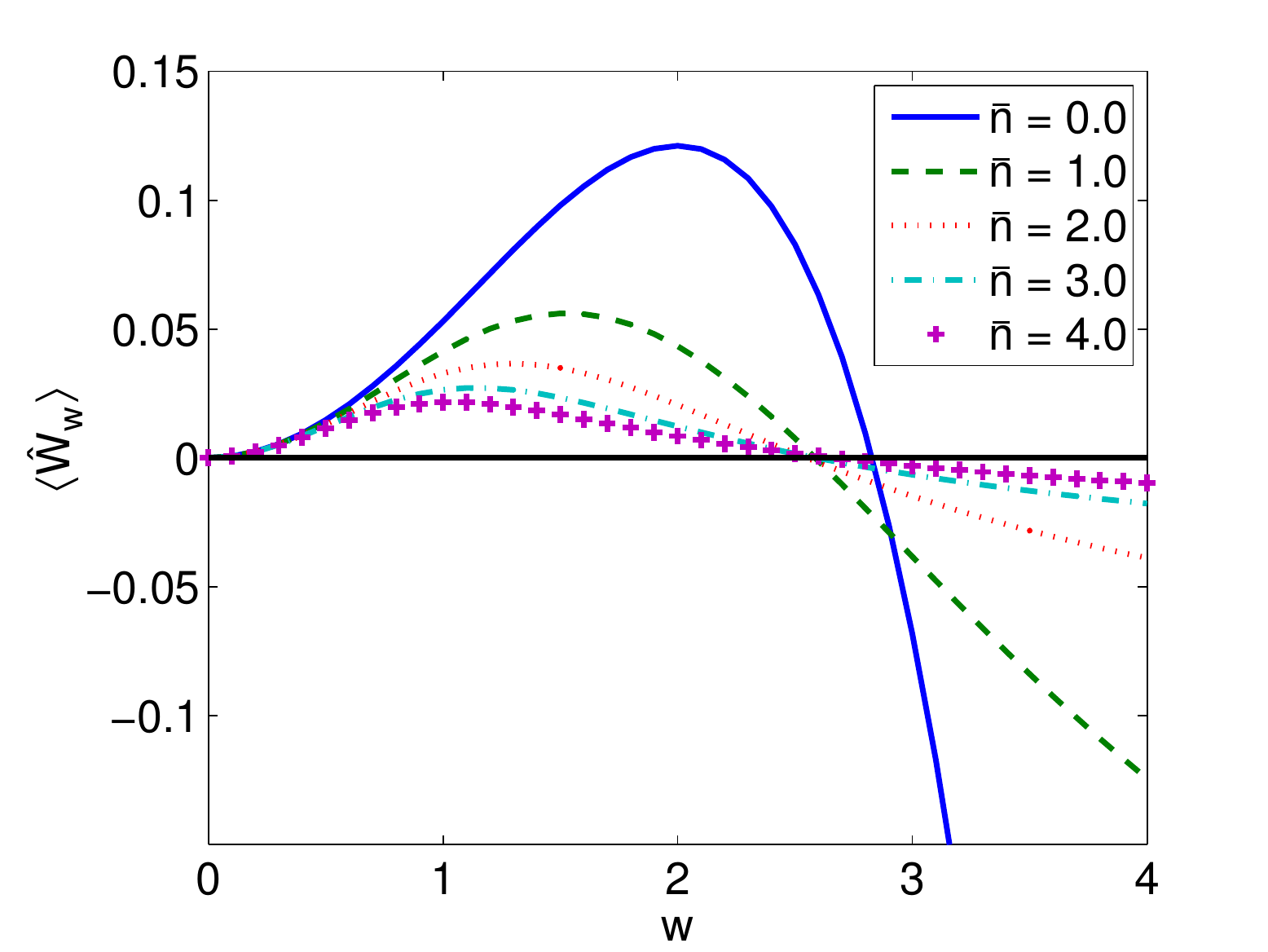}
   \caption{(Color online.) Dependence of the expectation value of the 
witness $\hat W_\w(0)$ on the filter width, for SPATSs of different mean 
thermal photon number $\bar n$ and quantum efficiency $\eta = 0.5$. 
Nonclassicality is indicated by negative expectation values, which 
appear for sufficiently large $\w$ for all states.}
   \label{fig:P:Omega:0}
\end{figure}

For our examinations, we choose the nonclassicality filter given in 
Eq.~\eqref{eq:triangular:witness}. Since SPATS are phase symmetric and 
similar to the photon, we may test the witness at the origin of phase 
space, i.e.~$\alpha = 0$. Figure~\ref{fig:P:Omega:0} shows the 
dependence of the expectation value of the nonclassicality witness $\hat 
W_\w$ on the filter width $\w$ for states with different mean thermal 
photon number $\bar n$. It is clearly seen that for sufficiently large 
$\w$, nonclassicality of all states is verified by negativity of the 
expectation value of the witness. The lower the mean thermal photon 
number, i.e.~the closer the state is to the photon, the more pronounced 
is the effect.

\section{Conclusions}

We provide universal nonclassicality witnesses, which identify the quantum effects of any quantum state of the harmonic oscillator. Our witness operators are controlled by only three real parameters, compared to the dependence of previously known witnesses on an infinite number of parameters. 
The detection of a witness requires a controlled coherent displacement of the quantum state and the determination of the photon number statistics. We demonstrate the
witnessing of a photon on a thermal background, which works very well even when other methods fail.

\section*{Acknowledgments} 
This work was supported by the Deutsche Forschungsgemeinschaft through SFB 652.

\appendix

\section{Construction of NFPs from witnesses.}
In the main article, we have already shown that any nonclassicality witness $\hat W$ with ${\rm Tr}(\hat W) = \frac{1}{\pi}$ can be used to define a nonclassicality filter $\Omega_1(\beta)$. Here, we construct a complete family of nonclassicality filters $\Omega_\w(\beta)$, which can be used for a complete nonclassicality test. In order to do so, we have to find a filter, equipped with a parameter $\w$, and with the following requirements:
\begin{enumerate}
	\item[(a)] The properties C1-C3 for a nonclassicality filter are satisfied.
	\item[(b)] For $\w = 1$, we want to return to the nonclassicality witness,  $\Omega_1(\beta) = \Phi^{(Q)}_{\hat W}(\beta)$.  
\end{enumerate}
We have already seen  that $\Phi^{(Q)}_{\hat W}(\beta)$ fulfills two conditions for a nonclassicality filter: The corresponding NFP is always finite for all $\alpha$, as we required by condition C1, and its negativities clearly indicate nonclassical effects, see condition~C2.  

The task is to introduce a width parameter $\w$ to the function $\Phi^{(Q)}_{\hat W}(\beta)$, such that the constructed family of functions satisfies all properties of a nonclassicality filter. For this purpose, we need an additional nonclassicality filter $\Omega'_\w(\beta)$, which can be constructed by an autocorrelation function as proposed in~\cite{Kiesel10}. Now, we look for suitable functions $f(\w), g(\w)$ such that 
\begin{equation}
	\Omega_\w(\beta) = f(\w) \Phi^{(Q)}_{\hat W} \left(\tfrac{\beta}{\w}\right) e^{-g(\w) |\beta|^2/2} + (1 - f(\w)) \Omega'_\w(\beta)\label{eq:construct:filter:from:W}
\end{equation}
is a nonclassicality filter and $\Omega_1(\beta) = \Phi^{(Q)}_{\hat W}(\beta)$.

The function $f(\w)$ shall be a weight function, i.e.~$0\leq f(\w)\leq 1$. Furthermore, for $\w = 1$, we want to return to the original nonclassicality witness, cf.~condition (b). This requires $f(1) = 1$ and $g(1) = 0$. Therefore, the function $f(\w) = \exp(-(\w-1)^2)$ is an appropriate choice. 

In order to have a nonnegative Fourier transform of $\Omega_\w(\beta)$, it is sufficient that each summand in Eq.~(\ref{eq:construct:filter:from:W}) has this property. Since $\Omega'_\w(\beta)$ is a nonclassicality filter itself, the statement is clear for the second term. Furthermore, the Fourier transform of $\Phi^{(Q)}_{\hat W}(\beta)$ is nonnegative, since
\begin{equation}
	\frac{1}{\pi^2} \int \Phi^{(Q)}_{\hat W}(\beta) e^{\alpha\beta^*-\alpha^*\beta} d^2\beta = \bra\alpha\hat W\ket\alpha.
\end{equation}
equals to the expectation value of the nonclassicality witness for the classical coherent states.  Therefore, it is sufficient that the remaining factor $e^{-g(\w) |\beta|^2/2}$ has a nonnegative Fourier transform, which can be achieved by requiring $g(\w) \geq 0$. 

Last, we require that $\Omega_\w(\beta) e^{|\beta|^2/2}$ is integrable for any real $\w\geq 1$, see condition C1. While the second term in Eq.~\eqref{eq:construct:filter:from:W} fulfills this condition by its definition, the first term has to be examined more carefully. We already know that $\Phi^{(Q)}_{\hat W}(\beta) e^{|\beta|^2/2}$ is integrable, since $\Phi^{(Q)}_{\hat W}(\beta)$ fulfills condition C1. Therefore, we split the term as follows:
\begin{eqnarray}
	&&\Phi^{(Q)}_{\hat W}
 \left(\tfrac{\beta}{\w}\right)  e^{-g(\w) |\beta|^2/2} e^{|\beta|^2/2}\nonumber\\
		&&= \left[\Phi^{(Q)}_{\hat W} \left(\tfrac{\beta}{\w}\right) e^{|\beta|^2/(2\w^2)}\right] e^{(1-g(\w)-1/\w^2) |\beta|^2/2}. 
\end{eqnarray}
As the factor in square brackets is integrable, the remaining factor must be bounded by one, which means
$g(\w) \geq 1 - 1/\w^2$. Altogether, the function $g(\w) = \max\{1-1/\w^2,0\}$ is a good choice, such that $\Omega_\w(\beta)$ defines a nonclassicality filter. This concludes our construction.

\end{document}